\documentclass[aps,prl,preprint,groupedaddress,showpacs]{revtex4-1}

\usepackage{graphicx}
\usepackage{color}
\usepackage[mathlines]{lineno}
\usepackage{amsfonts}
\usepackage{amsmath}
 \usepackage{nccmath}
    \newenvironment{mpmatrix}{\begin{medsize}\begin{pmatrix}}%
    {\end{pmatrix}\end{medsize}}%
\usepackage[none]{hyphenat}
\usepackage{textcomp}
\usepackage{epstopdf}

\let\originalleft\left
\let\originalright\right

\renewcommand{\left}{\mathopen{}\mathclose\bgroup\originalleft}
\renewcommand{\right}{\aftergroup\egroup\originalright}
\newcommand{\abs} [1]{\ensuremath{\left|#1\right|}}

\newcommand{\bra}[1]{\ensuremath{\left\langle#1\right|}}
\newcommand{\ket}[1]{\ensuremath{\left|#1\right\rangle}}
\newcommand{\bracket}[2]{\ensuremath{\left\langle#1 \vphantom{#2}\right| \left. #2 \vphantom{#1}\right\rangle}}

\begin{document}

% Use the \preprint command to place your local institutional report
% number in the upper righthand corner of the title page in preprint mode.
% Multiple \preprint commands are allowed.
% Use the 'preprintnumbers' class option to override journal defaults
% to display numbers if necessary
%\preprint{}

%Title of paper
\title{ Single-photon orbital angular momentum qudit states in fiber \--- \\ Limits to Dephasing correction via dynamical decoupling}

% repeat the \author .. \affiliation  etc. as needed
% \email, \thanks, \homepage, \altaffiliation all apply to the current
% author. Explanatory text should go in the []'s, actual e-mail
% address or url should go in the {}'s for \email and \homepage.
% Please use the appropriate macro foreach each type of information

% \affiliation command applies to all authors since the last
% \affiliation command. The \affiliation command should follow the
% other information
% \affiliation can be followed by \email, \homepage, \thanks as well.
%\author{}
%\email[]{Your e-mail address}
%\homepage[]{Your web page}
%\thanks{}
%\altaffiliation{}
%\affiliation{}

\author{Manish K. Gupta}
\email{\textcolor{magenta}{mgupta3@lsu.edu}}
\affiliation{Hearne Institute for Theoretical Physics, Department of Physics and Astronomy, Louisiana State University, Baton Rouge, Louisiana 70803, USA.}
\author{Jonathan P. Dowling}
\affiliation{Hearne Institute for Theoretical Physics, Department of Physics and Astronomy, Louisiana State University, Baton Rouge, Louisiana 70803, USA.}
%Collaboration name if desired (requires use of superscriptaddress
%option in \documentclass). \noaffiliation is required (may also be
%used with the \author command).
%\collaboration can be followed by \email, \homepage, \thanks as well.
%\collaboration{}
%\noaffiliation

\date{\today}

\begin{abstract}
We analytically derive a decoherence model for orbital angular momentum states of a photon in a multimode optical fiber and show that rate of decoherence scales exponentially with $l^2$, where $l$ is the azimuthal mode order.~We also show numerically that for large values of $l$ the orbital angular momentum photon state completely dephases.~However for lower values of $l$ the decoherence can be minimized by using dynamical decoupling to allow for qudit high-bandwidth quantum communication and similar applications.
\end{abstract}

% insert suggested PACS numbers in braces on next line
% insert suggested keywords - APS authors don't need to do this
%\keywords{}

%\maketitle must follow title, authors, abstract, \pacs, and \keywords
\maketitle

For the past few years the quantum information community has been putting a great deal of effort into boosting the bit rate for photonic quantum state transmission by encoding more than one bit per photon.~This is done by exploiting multiple temporal, spatial, polarization, and frequency modes of the single photon and then preparing a single photon in a superposition of those modes as a qudit.~The number of bits then is $\log_{2}d$, where $d$ is the dimension of the qudit.~A great deal of focus has been on using orbital angular momentum (OAM) modes of the photon, particularly in multimode optical fiber, as a road to high bit rate.

Photons that are OAM eigenstates, originate as a consequence of spatial distribution of optical field intensity and phase \cite{OAMBook2003,SLA2008}.~The photon beams carry an azimuthal phase term $\exp \left( i l \theta \right)$ and $l$ units orbital angular momentum per photon \cite{PhysRevA.45.8185}.~Such phase dependence is characteristic of either Laguerre-Gaussian or Bessel modes and each of these mode families provides a higher dimensional state space.~The most immediate advantage of large state space is large alphabet size for quantum communication and hence considerable increase in data capacity.~Higher dimensional systems have been known to improve security in quantum cryptography \cite{PhysRevLett.85.3313} and are required by some quantum network protocols \cite{PhysRevA.65.042320} and quantum computation schemes \cite{PhysRevA.65.052316} to efficiently solve problems like Byzantine agreement problem \cite{PhysRevLett.87.217901} and quantum coin tossing \cite{Ambainis2004398}.    

There are several protocols that encode quantum information in the two-dimensional Hilbert space of the photon's spin and exploit the polarization or time-bin degrees of freedom \cite{BB84, PhysRevLett.67.661}.~Physical implementations of one such protocol for quantum key distribution has shown that such a encoding is not optimal for practical applications due to a low bit rate \cite{1MbQKD2010}.~Information encoding based on the two-dimensional Hilbert space of photon polarization (or SAM) imposes a limitation on the rate of optical communication.~To overcome such limitations the orbital angular momentum (OAM) of light has been proposed that uses the photon's spatial mode structure and allows use of higher-dimensional Hilbert space, or a ``qudit" encoding of a photon \cite{16DQKD2013}.~This leads to an increased alphabet size and subsequently, increased rate of communication \cite{PhysRevLett.88.013601, PhysRevA.61.062308, keyZeilinger2006}.~Recent experiments have shown that the classical data-carrying capacity of a terabit per second can be achieved using OAM states of light in an optical fiber \cite{Bozinovic28062013}.~The potential of higher dimensional encoding of quantum information to achieve a higher bit rate can only be achieved if the photon can be protected from the decohering effect of optical index of refraction fluctuation in an optical fiber.

Here we report using a detailed calculation; an analytical model for decoherence caused by the refractive index fluctuation in a multi-mode fiber for an OAM photon state.~We show that rate of decoherence is faster for large values of $l$ and it scales exponentially with $l^2$, where $l$ is azimuthal mode number.~We additionally show that such a decoherence can be mitigated to a large extent with a \emph{open-loop} control technique called dynamical decoupling (DD) and we numerically show that OAM photon with small values of $l$ (up to about $10$) can be preserved with a fidelity greater than $99 \%$.

The transverse spatial wave function of a paraxial beam is an eigenstate of OAM  and it can be written in cylindrical coordinates as %
\begin{align} \label{eq:spatial_OAM}
\varphi_{pl}\left(r,\theta \right) = \frac{1}{\sqrt{2\pi}}  R_{p,l}\left( r \right) \exp{ \left(i \, l \, \theta \right)} .
\end{align}
The functions $R_{p,l}\left( r \right) $  are a basis for the radial dependence, such as the Laguerre-Gauss functions.~They are  defined as
\begin{align}
 R_{p,l}\left( r \right)  &=\frac{A}{w\left( z \right)}\left( \frac{\sqrt{2} r}{w\left(z\right)}\right)^{\abs{l}} L_{p}^{\abs{l}} \left( \frac{2{r}^2}{w\left(z\right)^2}\right)  \nonumber \\
 & \quad \times e^{ik{r}^2/\left[ 2R\left(z\right)\right]} \mathrm{e}^{i\left( 2p+\abs{l}+1 \right)\tan^{-1}{\left( z/z_{R} \right)}},
\end{align}%
where $w(z)=w_{0}\sqrt{1 + \left(z/z_{R} \right)^2}$ is the beam width, $R(z) = z\left[1 + (z_{R} /z)^2 \right]$ is the radius of wave-front curvature, and $z_{R} = \frac{1}{2} kw_{0}^2 $ is the Rayleigh range.~The quantity $\tan^{-1}(z/z_{R} )$ is known as the Gouy phase.

For simplicity, we consider an OAM photon that is launched in to a multimode optical fiber that is in superposition of $l$ and $-l$  states and has the following ket representation%
\begin{align}
\ket{\psi_{pl}}&= \frac{1}{\sqrt{2\pi}} R_{p,l}\left( r \right) \left[ \exp{ \left(i \, l \, \theta \right)} \ket{p,l} \right.  \nonumber \\ &\qquad \quad \left. +~\exp{\left(-i \, l \, \theta \right)}\ket{p,-l} \right].
\end{align}%
For example, such a state could be used as one code letter of a four-letter code for BB84 protocol \cite{BB84}.~The other letters would be the negative superposition and the individual $\pm l$ states.~The density matrix for the above input state can be written as 
%\begin{gather}
\begin{align}
\hat{\rho}_{\mathrm{in}} &= \ket{\psi_{pl}} \bra{\psi_{pl}}  \nonumber \\ &= \abs{R_{p,l}\left( r\right)}^{2}
\begin{pmatrix}
1 & \mathrm{e}^{i \, 2 l \theta} \\
\mathrm{e}^{-i \, 2 l \theta} & 1
\end{pmatrix} .
\end{align}
%\end{gather}

In general, the index of refraction fluctuation in an optical fiber can be represented by a series of concatenated, homogeneous segments of length $\Delta L$ with constant index fluctuation $\Delta \beta = \frac{\omega \left( n_{l}-n_{-l} \right) }{c}$ \cite{PhysRevA.91.032329,935820}.~When a photon that is in superposition of $+l$ and $-l$ propagates through the fiber in $z$ direction the $E$-fields see a slightly different refractive index due to the corkscrew nature of the OAM photon.~The two independent index of refraction fluctuations interact with the orbital angular momentum degree of freedom of the photon.~The noise operator is given by %
%\begin{gather}
\begin{align}
\mathbb{M}_{z}\left(\delta\phi_{j} \right) & = 
\begin{pmatrix}
\mathrm{e}^{i  \frac{\delta\phi_{j} }{2}} & 0 \\
0 & \mathrm{e}^{-i \frac{\delta\phi_{j} }{2}}
\end{pmatrix} \nonumber \\
&= \cos \left( \frac{\delta\phi_{j}}{2} \right) \mathbb{I} + i \sin \left( \frac{\delta\phi_{j} }{2} \right) \hat{L}_{z} \nonumber \\
&= \mathrm{e}^{i \frac{\delta\phi_{j} }{2} \hat{L}_{z}}  \nonumber \\ 
&= \mathbb{R}_{z}\left(\delta\phi_{j} \right)
\end{align}
%\end{gather}
where $\delta\phi_{j}  = \Delta \beta_{j} \Delta L$ is the phase angle acquired due to propagation through the $j^{th}$ segment of fiber and $\hat{L}_{z}=-i \frac{\partial}{ \partial \theta}$ is orbital angular momentum operator that generates rotation about $z$ axis.~Laguerre-Gaussian beams are eigenfunction of orbital angular momentum operator $\hat{L}_{z}$.~The output density matrix after the interaction in the $j^{th}$ segment is given by %
%\begin{gather}
\begin{align}
\hat{\rho}_{j_{\mathrm{out}}} &= \mathbb{M}_{z}\left(\delta\phi_{j} \right) \hat{\rho}_{in} \mathbb{M}_{z}\left(\delta\phi_{j} \right)^{\dagger} \nonumber \\
&= \abs{R_{p,l}\left( r \right)}^{2} 
\begin{mpmatrix}
1& \mathrm{e}^{i \, \left( 2\, l \, \theta \, + \,  \delta\phi_{j}\right)} \\
\mathrm{e}^{- i \, \left( 2 \, l \, \theta \, + \, \delta\phi_{j}\right)}  & 1
\end{mpmatrix} .
\end{align}
%\end{gather}

Now, if we assume that cross-talk between OAM modes is negligible, which is a good approximation for linear interactions, then the above density matrix can be rewritten as:
%\begin{gather}
\begin{align}
\hat{\rho}_{j_{\mathrm{out}}} = \abs{R_{p,l}\left( r \right)}^{2} 
\begin{mpmatrix}
1& \mathrm{e}^{i \,l \left( 2\, \theta \,+\,  \delta\phi_{j}\right)} \\
\mathrm{e}^{- i \, l \left( 2\, \theta \, + \, \delta\phi_{j}\right)}  & 1
\end{mpmatrix} .
\end{align}
%\end{gather}

After passing through the fiber with $n$ homogeneous concatenated segments the output density matrix is
%\begin{widetext}
\begin{align}\label{eq:densityMatrix}
\hat{\rho}_{j_{\mathrm{out}}} = \abs{R_{p,l}\left( r \right)}^{2}
\begin{mpmatrix}
1 &  \mathrm{e}^{i \,\left( 2 \, l \, \theta\right)} \, \prod \limits_{j=1}^n \,  \mathrm{e}^{ i \, \left( l \,\delta\phi_{j} \right)} \\
\mathrm{e}^{- i \, \left( 2 \, l \, \theta \right)} \, \prod \limits_{j=1}^n \,   \mathrm{e}^{ -i \,  \left( l \, \delta\phi_{j} \right) }  & 1
\end{mpmatrix} .
\end{align}
%\end{widetext}

We model the set of acquired phases $\left\lbrace \delta\phi_{1}, \delta\phi_{2},......,\delta\phi_{n} \right\rbrace$ as random variable  $\hat{\phi}$ with a mean $\left\langle \hat{\phi} \right\rangle = \phi_{0}$ and a nonzero variance $\left\langle \Delta\hat{\phi}^2 \right\rangle =  \Delta\phi^2 $.~The factor $\prod \limits_{j=1}^n \mathrm{e}^{\pm i  \, l \, \delta\phi_{i}}$ in the off-diagonal term of Eq. \ref{eq:densityMatrix} can be expressed in terms of mean and variance of random variable $\hat{\phi}$ %
\begin{align} \label{eq:phase}
\prod \limits_{j=1}^n \mathrm{e}^{\pm i  \, l \, \delta\phi_{i}} &= \exp \left[ \sum_{j=1}^n \left( \pm i  \, l \, \delta\phi_{i}\right) \right] \nonumber \\ 
&=\exp \left[ \sum_{j=1}^n \left( \pm i \, l \, \left\langle \hat{\phi} \right\rangle \pm  i \, l \, \Delta\hat{\phi} \right) \right]  \nonumber \\
&= \exp \left[ \pm i\, n \, l \, \left\langle \hat{\phi} \right\rangle \right] \exp \left[ \pm  i \, n \, l \, \Delta\hat{\phi} \right].
\end{align} 

We then Taylor expand the factor $\exp \left[ \pm  i \, n \,  l \, \Delta\hat{\phi} \right] $ of Eq. \ref{eq:phase} and take the time average to obtain
\begin{align} \label{eq:deco_factor_2}
\left\langle \exp \left[ \pm  i \, n \, l \,  \Delta\hat{\phi} \right] \right\rangle &= \left\langle 1 \pm i \, n \, l \,  \Delta\hat{\phi} - \frac{1}{2} n^2 \, l^2 \, \Delta\hat{\phi}^2 + \cdots \right\rangle \nonumber \\
&= 1 \pm i \, n \, l \, \left\langle \Delta\hat{\phi} \right\rangle - \frac{1}{2} n^2 \,  l^2 \, \left\langle \Delta\hat{\phi}^2 \right\rangle + \cdots
\end{align}%
Since the mean of variance is zero in Eq. \ref{eq:deco_factor_2}, and average of the variance is $\left\langle \Delta\hat{\phi}^2 \right\rangle =  \Delta\phi^2 $, hence we obtain the expression \cite{PhysRevA.91.032329, PhysRevA.88.023857}
\begin{align} \label{eq:decoherence_term}
\left\langle \exp \left[ \pm  i \, n \, l \, \Delta\hat{\phi} \right] \right\rangle &= 1 - \frac{1}{2} n^2 \, l^2 \,  \Delta\phi^2  + \cdots 
& \approx \mathrm{e}^{-\frac{1}{2} n^2 \, l^2 \, \Delta \phi^2} .
\end{align}%
Using Eq. \ref{eq:decoherence_term}, we can write Eq. \ref{eq:phase} as
\begin{align} \label{approximation}
\left\langle \prod \limits_{j=1}^n \mathrm{e}^{\pm i \, l \, \delta\phi_{i}} \right\rangle &= \left\langle \exp \left[ \pm i \, n \, l \, \left\langle \hat{\phi} \right\rangle \right]  \right\rangle \left\langle  \exp \left[ \pm  i \, n \, l \,  \Delta\hat{\phi} \right] \right\rangle \nonumber \\
&= \mathrm{e}^{ \pm i \, n \, l\, \phi_{0} } \, \mathrm{e}^{-\frac{1}{2} n^2 \, l^2 \, \Delta \phi^2}.
\end{align}%
And finally, with the expression obtained in Eq. \ref{approximation}, the output density matrix in Eq. \ref{eq:densityMatrix} can be rewritten as
\begin{widetext}
\begin{align} \label{eq:apprxDensityMatrix}
\hat{\rho}_{\mathrm{out}} = \abs{R_{p,l}\left( r \right)}^{2}
\begin{pmatrix}
1 & \mathrm{e}^{ i \, l \, \left( 2 \, \theta \, + \, n \, \phi_{0}\right)}  \mathrm{e}^{-\frac{1}{2} n^2 \, l^2 \, \Delta \phi^2}\\
\mathrm{e}^{ -i \, l \, \left( 2 \, \theta \, + \, n \, \phi_{0}\right)} \mathrm{e}^{-\frac{1}{2} n^2 \, l^2 \, \Delta \phi^2} & 1
\end{pmatrix} .
\end{align}
\end{widetext}%
Here $n$ is a constant proportional to the total distance $z$ propagated through the fiber.~That is, without loss of generality, $n=kz$ where $k=2\pi/\lambda$.

The state represented by $\hat{\rho}_{\mathrm{out}}$ is no longer pure due to presence of the dephasing term $ \mathrm{e}^{-\frac{1}{2} n^2 \, l^2 \, \Delta \phi^2} $ in the off-diagonal terms and the rate of decoherence is much faster for larger values of $l$. 

To understand the detrimental effects of noise encountered in the communication channel we numerically study three scenarios.~First we analyze the decoherence of free evolving OAM photon in a fiber due to index of refraction fluctuations and then we now analyze the effectiveness of open-loop control in preserving the coherence of the qubit, where the system is subjected to external, suitably tailored, space-dependent pulses which do not require measurement.~Finally we analyze the impact of large values of quantum number $l$ on decoherence suppression.

Decoherence of a photonic state has its origin in optical index fluctuation of a fiber that can result from both intrinsic and extrinsic perturbations.~We model axially varying index dephasing in an optical fiber of length $L$ by a series of concatenated, homogeneous segments of length $\Delta L$ with constant $\Delta n$ \cite{PhysRevA.91.032329,935820,PhysRevA.85.022340}.~The index fluctuations across these segments is modeled by generating a set of values according to the Rayleigh distribution, whose probability density function is given as%
\begin{equation}
f(x,\sigma) = \frac{x}{\sigma^2}e^{{-x^2}/{(2 \sigma^2)}},~x \geq 0
\end{equation}
where $\sigma \geq 0$, is the scale parameter of the distribution, and $x$ is the distance along the fiber \cite{Rayleigh,935820}.~A noise profile of the fiber is extrapolated from these phase error values.~Here we assume that the fiber only exhibits linear index fluctuation as the radial dimension of the fiber is very small.

 For our numerical analysis we consider the following input state

\begin{align}\label{eq:inputState}
\ket{\psi} &= \frac{1}{\sqrt{2\pi}} \left[ R_{p,l}\left( r \right) \exp{ \left(i \, l \, \theta\right)} \ket{p,l} \right. \nonumber \\
& \qquad \quad \left. + R_{p,-l}\left( r \right) \exp{\left(-i \, l \, \theta \right)}\ket{p,-l} \right].
\end{align}%
Since $R(p,l) =R(p,-l)$, we can normalize the state and rewrite the ket in matrix notation as:
\begin{equation}\label{eq:ket_OAM}
\ket{\psi} = \frac{1}{\sqrt{2}}
\begin{pmatrix}
\mathrm{e}^{i \,  l \phi} \\
\mathrm{e}^{-i \,  l \phi} 
\end{pmatrix} .
\end{equation}

We first calculate the fidelity of the fiber without any error suppression mechanism in place for a particular length, number of sections, and initial state.~The initial state of the photon is allowed to freely evolve through each section of the fiber according to %
\begin{equation} \label{eq:free_evol}
\mathbb{M}_{z}\left(\delta\phi_{j}\right)  = 
\begin{pmatrix}
\mathrm{e}^{i \frac{\delta\phi_{j}}{2}} & 0 \\
0 & \mathrm{e}^{-i \frac{\delta\phi_{j}}{2}}
\end{pmatrix},
\end{equation}
where, $\delta\phi_{j}$ includes the phase error from the Rayleigh distribution.~The freely evolved state is then compared with the input state.~We use fidelity as a measure of effectiveness in preserving the state of photon and it is defined as%
\begin{equation}
\mathcal{F}= \abs{\bracket{\psi_{i}}{\psi_{f}}}^{2} = \bra{\psi_{i}} \widehat{\rho}_{\mathrm{out}} \ket{\psi_{i}},
\end{equation}
where $\psi_{f}$ and $\psi_{i}$ represent the final and initial state respectively and $\widehat{\rho}_{\mathrm{out}}= \frac{1}{n}\sum{ \ket{\psi_{out}} \bra{\psi_{out}}}$ is average output state over $n$ fiber noise profiles.

\begin{figure}[ht]
\includegraphics[width=11cm]{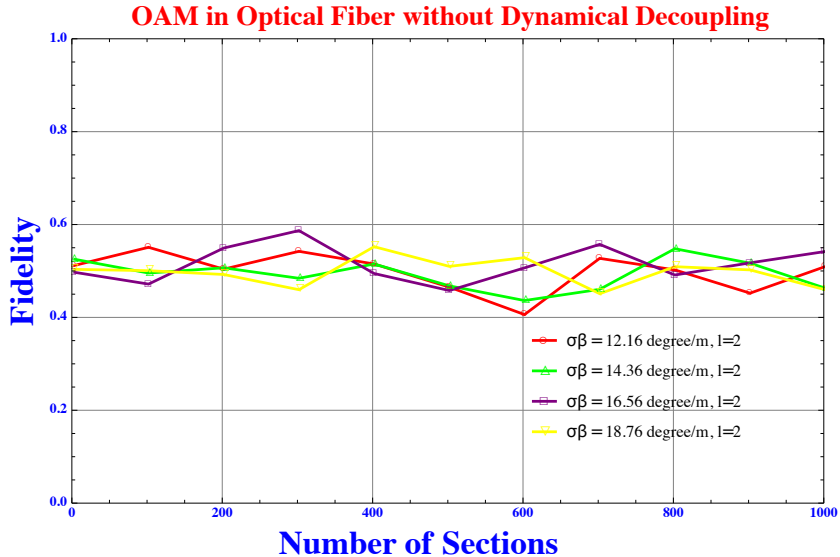}% Here is how to import EPS art
\caption{\label{fig:OAM_SMF_NO_DD} (Color Online) Fidelity of an OAM state in a $500$ m optical fiber without dynamical decoupling. The value of $0.5$ indicated a maximally mixed state.}
\end{figure}

When photon with state of the form Eq. \ref{eq:ket_OAM} is launched into the fiber of length $500$ m it completely dephases as shown in the Fig. \ref{fig:OAM_SMF_NO_DD} and fidelity remains at $50\%$.

\begin{figure}[ht]
\includegraphics[width=11cm]{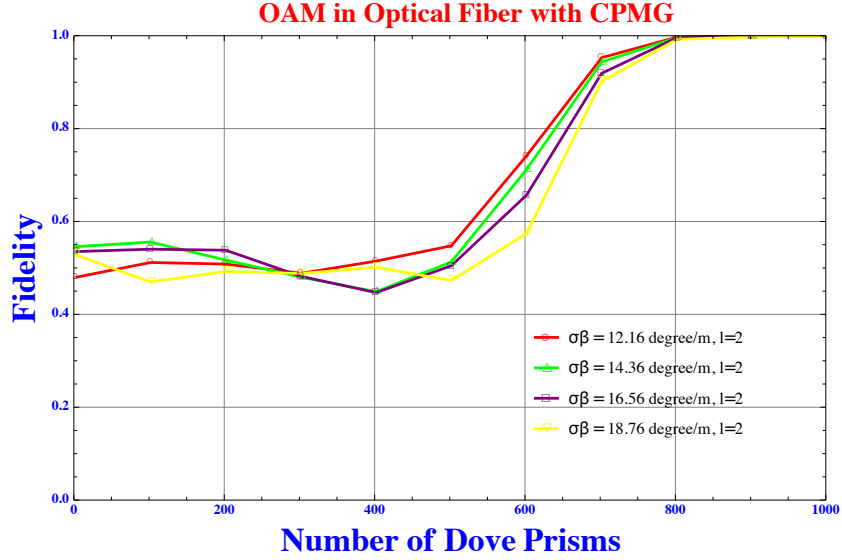}% Here is how to import EPS art
\caption{\label{fig:OAM_SMF_CPMG_l_2} (Color Online) Fidelity of CPMG sequence in a $500$ m optical fiber with perfect pulses.~The result shown in the plot is for an OAM state with arbitrary $\phi$ and $l=2$.}
\end{figure}

We then calculate fidelity for second scenario where the passive error suppression mechanism called the Carr-Purcell-Meiboom-Gill (CPMG) DD pulse sequence is used for a particular length, number of sections, and initial state.~For each section of fiber, the initial state of photon is allowed to freely evolve for a certain distance according to Eq.~\ref{eq:free_evol} and then the state is  rotated according to the CPMG DD pulse sequence, where the pulse sequence is implemented by inserting a dove prism.~This prism is a well-known device in optics that acts as image flipper in one transverse dimension, while leaving unchanged the image in the other transverse dimension \cite{Gonzalez:06}.~This changes the OAM of a light beam from $l=1$ to $l=-1$.~This is repeated for each section in the fiber.~We then compare the output state with input state and use fidelity as a measure of effectiveness in preserving the state of photon.~We see that for $l=2$ the photon state can be preserved with a fidelity greater than $99\%$ as shown in Fig. \ref{fig:OAM_SMF_CPMG_l_2}.

\begin{figure}[ht]
\includegraphics[width=11cm]{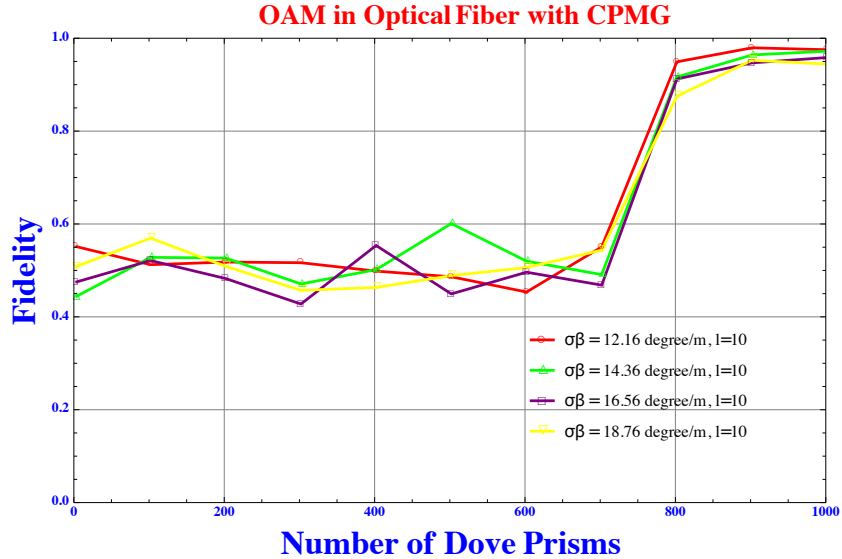}% Here is how to import EPS art
\caption{\label{fig:OAM_SMF_CPMG_l_10} (Color Online) Fidelity of CPMG sequence in a $500$ m optical fiber with perfect pulses.~The result shown in the plot is for an OAM state with arbitrary $\phi$ and $l=10$.}
\end{figure}

\begin{figure}[ht]
\includegraphics[width=11cm]{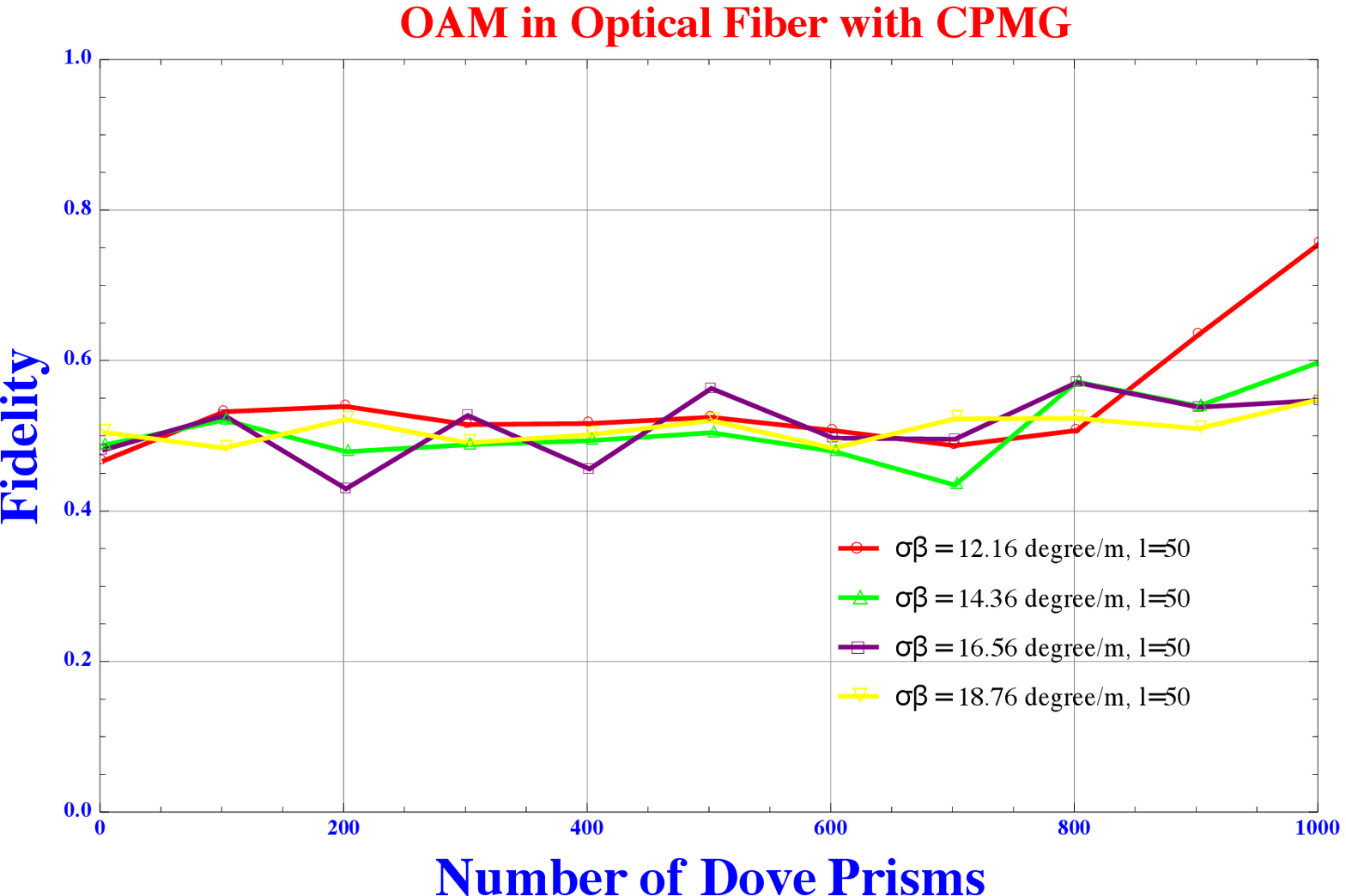}% Here is how to import EPS art
\caption{\label{fig:OAM_SMF_CPMG_l_50} (Color Online) Fidelity of CPMG sequence in a $500$ m optical fiber with perfect pulses.~The result shown in the plot is for an OAM state with arbitrary $\phi$ and $l=50$.}
\end{figure}

\begin{figure}[ht]
\includegraphics[width=11cm]{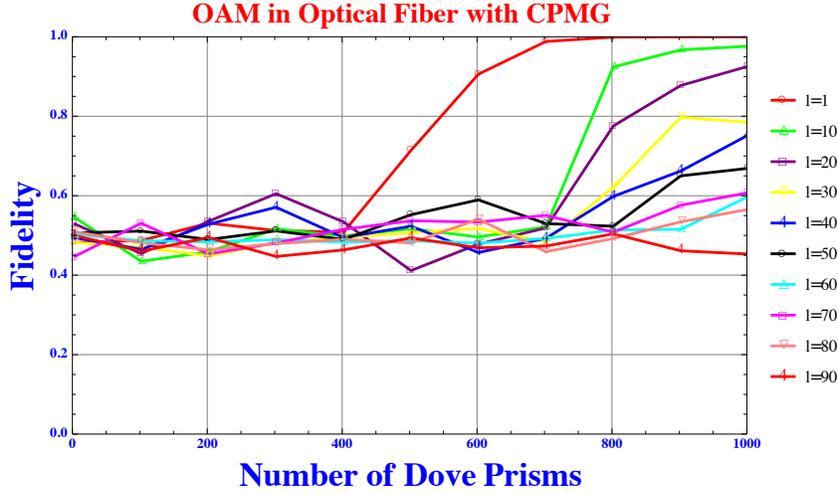}% Here is how to import EPS art
\caption{\label{fig:OAM_SMF_CPMG_l_1_100} (Color Online) Fidelity of CPMG sequence in a $500$ m optical fiber with perfect pulses.~The result shown in the plot is for an OAM state with arbitrary $\phi$ and $l$ between one and $100$.}
\end{figure}

Finally we analyze the impact of large quantum number $l$ on effectiveness of CPMG DD pulse sequence in preserving the OAM state of photon.~We find that the fidelity decreases for same number of resource with increasing value of quantum number $l$ such as $l=10$ and $l=50$ as shown in Fig. \ref{fig:OAM_SMF_CPMG_l_10} and Fig. \ref{fig:OAM_SMF_CPMG_l_50}.~As $l$ is increased from $1$ to $100$, we see that the DD pulse sequence fails to preserve the OAM state of photon and it completely  dephases as shown in the Fig. \ref{fig:OAM_SMF_CPMG_l_1_100}

We have derived the decoherence model for OAM transport in a optical fiber and have shown that the rate of decoherence is dependent on $l^2$.~We also show numerically that the OAM state can be preserved against decoherence caused by the index fluctuation present in a  fiber with $>99\%$ fidelity using CPMG dynamical decoupling scheme up to a certain maximum value of $l$.~For quantum optimal communication schemes, such as quantum key distribution, one would like to put a single-photon into a superposition state of the highest possible number of OAM states.~That is because the number of bits per photon scales as $\rm{log}_{2}\left( d \right)$, where $d$ is the dimension of the qudit.~For example, encoding in a superposition of up to $p$ and $l$ OAM states gives%
\[
d = 2 \frac{\left(p \left( p+1 \right)\right)}{2} \frac{\left(l \left( l+1 \right)\right)}{2}.
\]
Our work here indicates that dephasing will limit $l_{\rm{max}}$ to $l=10$ for most scenarios.~Dephasing cannot be corrected by DD beyond that value of $l$. 

A state of the form Eq. \ref{eq:inputState}, which is an equal superposition of OAM $l$ and $-l$, can be prepared by starting with a linear polarized light at $45$\textdegree~and Mach \textendash Zehnder interferometer with quarter-wave plate in both the arms and Dove prism in one of the arms.

% If you have acknowledgments, this puts in the proper section head.
\begin{acknowledgments}
The authors would like to acknowledge support from the Air Force Office of Scientific Research, the Army Research Office, the National Science Foundation and the Northrop Grumman Corporation.
\end{acknowledgments}

% Specify following sections are appendices. Use \appendix* if there
% only one appendix.
%\appendix
%\section{}

% Create the reference section using BibTeX:
\bibliography{References}

\end{document}